\documentclass{svjour3}                     
\smartqed  
\usepackage{graphicx}
\journalname{Ocean Dynamics}
\begin{document}

\title{Eddy energy sources and mesoscale eddies in the Sea of Okhotsk.}

\titlerunning{EKE IN THE OKHOTSK SEA}  

\author{D.V. Stepanov}

\authorrunning{Stepanov} 

\institute{D. V. Stepanov \at
              Department of the Ocean and Atmosphere Physics, \\
              V.I. Ilichev Pacific Oceanological Institute, Vladivostok, Russia \\
              Tel.: +7423 231-1400\\
              Fax:  +7423 231-2573\\
              \email{step-nov@poi.dvo.ru}
}

\date{Received: date / Accepted: date}

\maketitle

\begin{abstract}
Based on an eddy-permitting ocean circulation model, the eddy
kinetic energy (EKE) sources are studied in the Sea of Okhotsk. An
analysis of the spatial distribution of the EKE showed that intense
mesoscale variability occurs along the western boundary of the Sea
of Okhotsk, where the East-Sakhalin Current extends. It was found a
pronounced seasonally varying EKE with its maximal magnitudes in
winter, and its minimal magnitudes in summer.

An analysis of the EKE sources and the energy conversions showed
that time-varying (turbulent) wind stress is a main contribution to
mesoscale variability along the western boundary of the Sea of
Okhotsk. The contribution of baroclinic instability to the
generation of mesoscale variability  predominates over that of
barotropic instability along the western boundary of the Sea of
Okhotsk.

To demonstrate the mechanism of baroclinic instability, the
circulation was considered along the western boundary of the Sea of
Okhotsk from January to April 2005. An analysis of hydrological
conditions showed outcropping isopycnals and being strong vertical
shear of the along-shore velocity from January to May 2005. In
April, mesoscale eddies are observed along the western boundary of
the Sea of Okhotsk. It was established that seasonal variability of
turbulent wind stress and the baroclinic instability of the
East-Sakhalin Current are major reasons of mesoscale variability
along the western boundary of the Sea of Okhotsk.


\end{abstract}

\section{Introduction}
\label{intro}

The Sea of Okhotsk is one of the marginal seas in the north-western
Pacific Ocean. This sea is situated at high latitudes and being the
southernmost sea, covered by sea ice during the year. The sea ice
covering period begins from the end of November and can last up to
the end of May. It complicates significantly to carry out
observations in the Sea of Okhotsk.

The Okhotsk Sea circulation is the subject of high scientific
interest \cite{Ohshima2002,Mizuta2003}, because this basin is the
source of intermediate water in the western North Pacific
\cite{Talley1991,Gladyshev2003,Fukamachi2004}. Dense shelf waters,
generated over the north-west shelf of the Sea of Okhotsk in winter,
are transported by the East-Sakhalin Current (ESC) in the Kuril
Basin, where they are mixed by mesoscale eddies and tides. Scherbina
et al. ~\cite{Shcherbina2004a}, based on datasets obtained from the
moorings deployed at the north-western this sea, discovered sharp
changes in the density increase in late February. Authors supposed
that these sharp density changes were induced by baroclinic
instability of the density front.

The basic element of the basin-scale circulation in the Sea of
Okhotsk is the ESC, which extends from the north-western to the
southern part of this sea. In the south-western Sea of Okhotsk, the
Kuril Basin is situated with the depth exceeding 3000 m. The
anticyclonic circulation occurs over the Kuril Basin. The Sea of
Okhotsk interacts with the Japan/East Sea by means of the Soy and
Tatar Straits and with the North Pacific Ocean by means of the Kuril
Straits.

Numerous studies investigated reasons and mechanisms of the
basin-scale circulation in the Sea of Okhotsk. According to
\cite{Ohshima2002,Smizu2006}, it is supposed that the wind stress is
the major driver of the basin-scale circulation in the Sea of
Okhotsk. The dominating positive wind stress curl over the Sea of
Okhotsk drives cyclonic circulation in the central part of this sea.
In addition, the along-shore wind stress component is the major
driver of the along--shore branch of the ESC. Because of strong
seasonal variability of the wind stress, the intensity of the
Okhotsk Sea circulation also exhibits strong seasonal variability.
In addition, heat and freshwater fluxes over the Sea of Okhotsk
exhibit seasonal variability and their impact on the basin-scale
circulation is corrected by ice covering.

Mesoscale eddies in the Sea of Okhotsk are the subject of intensive
investigations. Ohshima et al.~\cite{Ohshima2002}, based on
satellite-tracked drifter observations, revealed that anticyclonic
eddies with the diameter varying from 100 to 200 km dominate over
the Kuril Basin and eddy kinetic energy exceeds mean kinetic energy
from 3 to 20 times. Ohshima et al.~\cite{Ohshima2005} established
that the major mechanism of eddy generation over the Kuril Basin is
the baroclinic instability of the tidal front induced by intense
tidal mixing near the Kuril Straits. Ohshima et
al.~\cite{Ohshima1990} investigated mesoscale variability over the
south-western Kuril Basin near the Soya Strait. It was established
that the mechanism of eddy generation is the barotropic instability
of the Soya Current. The impact of the Soya Current transport on
mesoscale variability was studied by \cite{Uchimoto2007}. Thermal
infrared images with very high spatial resolution showed features of
sub-mesoscale variability (the eddy diameter ranging from 2 to 30
km) near the Kuril Islands \cite{Nakamura2012}.

In the above mentioned studies, mesoscale variability in the Sea of
Okhotsk was studied mainly during the ice-free period. Thus, the
whole picture of mesoscale variability in the Sea of Okhotsk does
not fully understand. Because of challenge of the carrying out
natural observations during the sea ice covering period, at studying
the Okhotsk Sea circulation, the numerical models are applied, which
are accounting for mesoscale variability. It should be noted that
according to the work \cite{Chelton1998}, at high latitudes the
first baroclinic Rossby radius of deformation $(\lambda_1)$ is
significantly less than that at middle and low latitudes. Low values
of $\lambda_1$ constrain spatial resolution of the model grid, which
needs to explicitly resolve mesoscale variability in the Sea of
Okhotsk. In addition, $\lambda_1$ magnitudes can exhibit significant
spatial and time variability due to strong irregularity of the
bottom topography and seasonal variability of density stratification
in the Sea of Okhotsk. Numerical simulations of the circulation in
the Sea of Okhotsk with high spatial resolution were presented in a
number of works. Based on numerical simulations with the grid
resolution of 3 km, Matsuda et al.~\cite{Matsuda2015,Matsuda2009}
analyzed ventilation processes in the intermediate layer of the Sea
of Okhotsk. However, reasons and mechanisms of mesoscale variability
in the Sea of Okhotsk have not fully revealed.

At investigating mesoscale variability both the World Ocean and
marginal seas, the one of the approaches is based on the analysis of
the eddy kinetic energy (EKE) budget. von Storch et
al.~\cite{Storch2012} presented the methodology and carried out the
comprehensive analysis of the sources and sinks of the EKE in the
World Ocean based on numerical simulations. They assessed
contributions of baroclinic and barotropic instabilities of the
large-scale currents to the generation of mesoscale variability.

Based on this methodology, the EKE budget and mechanisms of
mesoscale variability have examined in the Red Sea, the South China
Sea and the Labrador Sea. Based on the outputs of the
high-resolution MITgcm, Zhan et al.~\cite{Zhan2016} established that
the leading mechanism of the eddy generation in the Red Sea is
baroclinic instability, whereas turbulent wind stress and barotropic
instability influence weakly on mesoscale variability in this sea.
Based on the outputs of the high-resolution LICOM, Yang et
al.~\cite{Yang2013} established that both hydrodynamic instability
and wind power input influence significantly on mesoscale
variability in the South China Sea. Eden et al.~\cite{Eden2002},
based on the outputs of the MOM, established that the barotropic
instability of the West Greenland Current is the major source of EKE
in the Labrador Sea. Thus, depending on the considered basin,
contributions of wind power input, baroclinic instability and
barotropic instability to EKE can be different \cite{Storch2012}.

In this study, based on the numerical simulations, mesoscale
variability is analyzed in the Sea of Okhotsk. Wind power input,
baroclinic and barotropic instabilities of the ESC are considered as
the main sources of the EKE. The paper is organized as follows.
Sect.~\ref{sec:2} describes the model configuration and validation
of its outputs. Spatial and temporal analysis of EKE in the Sea of
Okhotsk is presented in Sect.~\ref{sec:3}. An analysis of eddy
energy conversion from the mean circulation is presented in
Sect.~\ref{sec:4}. Sect.~\ref{sec:5} presents estimations of the
main sources of EKE in the Sea of Okhotsk. Typical picture of
mesoscale variability on the eastern shelf of Sakhalin Island
induced by baroclinic instability of the ESC is presented in
Sect.~\ref{sec:6}. Discussion and summary of the main results are
presented in Sect.~\ref{sec:7}.

\section{Model setup and validation}
\label{sec:2}

To simulate the circulation in the Sea of Okhotsk, an INMOM model is
applied with the horizontal resolution of about 3.5 km and 35
sigma-levels compressed toward the sea surface to resolve density
stratification. The INMOM is the sigma-coordinate model based on
primitive equations of ocean dynamics with hydrostatic and
Boussinesq approximations \cite{Gusev2014,Stepanov2014,Diansky2016}.
The model domain covers the Sea of Okhotsk, the Japan/East Sea and
the north-western Pacific Ocean to take into consideration the water
exchange between of them. To obtain quasi-uniform spatial
resolution, the spherical coordinate system with the pole, situated
at the point with the coordinates of (25.5$^{\circ}$E,
22.4$^{\circ}$N), is used. Thus, an equator of a new coordinate
system crosses the Sea of Okhotsk and the Japan/East Sea.

To account for mesoscale variability in the Sea of Okhotsk, it is
necessarily that the spatial scale of the model grid and $\lambda_1$
are the same order. Preliminarily, $\lambda_1$ was assessed with the
relationship \cite{Chelton1998},
\begin{equation}
\label{eq1} \lambda_1 = \frac{c_1}{|f|},
\end{equation}
where $f=2\Omega\sin\vartheta$ is the Coriolis parameter, $\Omega$
is the Earth rotation rate, $\vartheta$ is the latitude and $c_1$ is
the first eigenvalue, which satisfies the boundary value problem
\begin{equation}
\label{eq2}
\begin{array}{l}
\frac{d^2}{dz^2}\phi_1+\frac{N^2(z)}{c_1^2}\phi_1=0,\\
\phi_1(0)=\phi_1(-H)=0.
\end{array}
\end{equation}
Here, the vertical coordinate $z$ directs from the center of the
Earth, $N(z)$ is the buoyancy frequency profile, $H$ is the depth
and $\phi_1(z)$ is the first eigenfunction of the boundary value
problem (\ref{eq2}). According to \cite{Chelton1998}, c$_1$ can be
assessed as
\begin{equation}
\label{eq3} c_1 \approx \frac{1}{\pi} \int_{-H}^0 N(\xi)d\xi.
\end{equation}
Based on climatological monthly mean temperature and salinity fields
of datasets \cite{Locarnini2013,Zweng2013}, $\lambda_1$ was
assessed. These fields have the horizontal resolution of
0.25$^{\circ}$ and to be arranged at 102 horizons.

Fig.~\ref{fig:1} shows an annual mean distribution of $\lambda_1$ in
the Sea of Okhotsk. According to the obtained estimations, maximum
magnitude of $\lambda_1$, amounting to 20 km, occurs in the
south-western Sea of Okhotsk over the Kuril Basin. Over the central
part of the Sea of Okhotsk, $\lambda_1$ magnitude varies from 10 to
12 km and from 3 to 9 km along the western boundary of this sea.
Minimal values of $\lambda_1$ occur in the north-eastern Sea of
Okhotsk and range from 1 to 2 km. Thus, the used horizontal
resolution is the eddy-permitting resolution except the
north-eastern and northern part of the Sea of Okhotsk, where the
spatial scale of the model grid is higher than $\lambda_1$.

Bottom topography of the model domain was extracted from the GEBCO
dataset \cite{Becker2009} and to be smoothed by the 9-point filter.
Sensible and latent heat fluxes, short- and long-wave radiation,
momentum flux and net salt flux, containing precipitation,
evaporation and climatological runoff contributions, are set with
the bulk-formulae \cite{Stepanov2014,Diansky2016,Large2009}.
Atmospheric parameters were extracted from the ERA-Interim dataset
\cite{Dee2011} with the spatial resolution of
0.75$^{\circ}\times$0.75$^{\circ}$ from 1979 to 2009. It should be
noted that wind velocity field at the high of 10 m, air temperature
and absolute humidity at the 2 m, as well as sea level pressure have
the time-resolution of 6 hours. To correctly account for the
interaction between the circulation in the Sea of Okhotsk and
atmospheric forcing, the INMOM includes a sea-ice model. This model
accounts for processes of generation and melting of sea ice and
transforming of stale snow to sea ice \cite{Yakovlev2003}. At the
same time, wind stress over the Sea of Okhotsk is calculated under
the \textit{open water} condition, that is, the sea ice covering is
absent. The approximation of the \textit{open water} is correct,
when the sea ice compactness is less than 0.75. When the sea ice
compactness is more than 0.75, then the approximation of the
\textit{open water} accounts for the impact of the moving ice on the
sea water.

Initial potential temperature and salinity are extracted from the
datasets \cite{Locarnini2013,Zweng2013}. On the sea surface,
potential temperature and salinity are corrected by adding their
climatological values to the heat and salinity fluxes with the
relaxation parameter amounts to 10/3 m month$^{-1}$. This procedure
removes the \textit{climatological drift} of the circulation in the
Sea of Okhotsk induced by uncertainties of atmospheric parameters.
It should be noted that the presented model configuration does not
account for the tidal impact on the circulation in the Sea of
Okhotsk.

On the open boundary of the model domain, no-normal and no-slip flow
conditions are set. In narrow regions near the open boundaries from
the sea surface to bottom, the nudging condition is set for
potential temperature and salinity with the relaxation parameter of
3 hours.

The sub-grid processes are parameterized with the viscosity operator
of the second order with the coefficient of 100 m$^2$ s$^{-1}$. The
horizontal diffusion of heat and salt, formulated along the
geopotential surfaces, are parameterized with the viscosity operator
of the second order with the coefficient of 10 m$^2$ s$^{-1}$ for
both variables. The vertical turbulent processes are parameterized
according to \cite{Pacanowski1981} and vertical viscosity and
diffusivity amount to 10$^{-4}$ m$^2$ s$^{-1}$ and 10$^{-5}$ m$^2$
s$^{-1}$, respectively. Convective mixing is parameterized by
maximum viscosity and diffusivity, which amount to
2.5$\times$10$^{-2}$ m$^2$ s$^{-1}$ and 5$\times$10$^{-3}$ m$^2$
s$^{-1}$, respectively.

Preliminarily, we simulated circulation during four years with the
atmospheric forcing corresponding to 1979. Initial conditions for
potential temperature and salinity corresponded to June. Thus, we
avoided setting up initial compactness and height of sea ice in our
model configuration. Model outputs, obtained in the end of fourth
year of numerical simulations, were used as initial conditions for
the numerical simulations with the atmospheric forcing varying from
1979 to 2009.

In this study, we analyze the model outputs from 2005 to 2009.
Fig.~\ref{fig:2} shows a long--term mean (from 2005 to 2009) of a
velocity field at the horizon of 10 m, as well as a mean of wind
power input
$(\overline{\mathbf{\tau}}\cdot\overline{\mathbf{u}_s})$, where
$\overline{\mathbf{\tau}},\overline{\mathbf{u}_s}$ are the wind
stress and sea surface currents averaged for the season,
respectively.

In winter, maximal velocities, varying from 0.3 to 0.4 m s$^{-1}$,
occur along the western boundary of the Sea of Okhotsk, where the
ESC extends. Along the norther boundary of the Sea of Okhotsk and
over the Kuril Basin, it is observed less intense currents with
velocities ranging from 0.1 to 0.15 m s$^{-1}$. In addition, maximal
wind power input, amounting to 4$\times$10$^{-2}$ W m$^{-2}$, occurs
over the western and northern boundaries of the Sea of Okhotsk.
Positive values of
$(\overline{\mathbf{\tau}}\cdot\overline{\mathbf{u}_s})$ indicate
that the current direction coincides with the direction of mean wind
stress, except a small region is situated near the eastern boundary
of the Sea of Okhotsk. From spring to summer, the simulated
circulation weakens significantly. Velocities in the ESC weaken from
0.2 m s$^{-1}$ to 0.1 m s$^{-1}$ and current direction in the
western of the Sea of Okhotsk changes opposite to direction of wind
stress, which weakens up to --0.2$\times$10$^{-2}$ W m$^{-2}$. In
autumn, the simulated circulation strengthens again in the northern
and the western Sea of Okhotsk up to 0.2--0.25 m s$^{-1}$. At the
same time, mean wind power input increases up to
2.5$\times$10$^{-2}$ W m$^{-2}$. It should be noted that
$(\overline{\mathbf{\tau}}\cdot\overline{\mathbf{u}_s})$ shows its
positive values in the regions of intense currents along the western
and northern boundaries of the Sea of Okhotsk. The simulated surface
circulation is similar that obtained from the natural observations
\cite{Moroshkin1966,Luchin1998} and the spatial distribution of mean
wind power input does not contradict conclusions about significant
influence of wind stress on the basin-scale circulation in the Sea
of Okhotsk.

Because the ESC is the most intense current in the structure of the
simulated circulation, we will concentrate on the velocity field
along the western boundary of the Sea of Okhotsk. The simulated
circulation shows that the ESC consists of two branches (cores). The
first branch of this current onsets in the north-western Sea of
Okhotsk. The second branch of the ESC is the element of the
basin-scale cyclonic gyre covering the central part of this sea. An
analysis of the simulated velocity field shows that the ESC exhibits
strong seasonal variability (not shown). At the horizon of 20 m,
monthly mean velocities in the first and second branches of the ESC
reach to 0.3 m s$^{-1}$ and 0.1 m s$^{-1}$, respectively. In spring,
velocities in the first branch of the ESC decrease up to 0.21 m
s$^{-1}$. In the end of summer, the intensity of the ESC is minimal.
On the eastern shelf of Sakhalin Island, monthly mean velocities are
limited by the value of 0.15 m s$^{-1}$. In autumn, the simulated
velocities in the ESC increase again up to 0.32 m s$^{-1}$.

The estimation of the annual mean ESC transport shows that in winter
it reaches the value of 6 Sv. In summer, the ESC transport shows its
minimal magnitudes ranging from 1.5 to 2 Sv. The obtained estimation
of the ESC transport and its season variability are similar with
that obtained from the natural observations \cite{Ohshima2002} and
numerical simulations of the Sea of Okhotsk circulation
\cite{Matsuda2015}. Comparing the results of the simulated
circulation with those of the previous studies shows that the mean
simulated circulation is correct and to be characterized by the
occurrence of the north current along the western boundary of the
Sea of Okhotsk. This north current consists of two branches and
features seasonal variability with maximum intensity in winter and
minimum intensity in late summer. The subject of next sections is
mesoscale variability and its main sources mainly along the western
boundary of the Sea of Okhotsk.

\section{Eddy kinetic energy in the Okhotsk Sea}
\label{sec:3}

In this section, based on the model outputs, the EKE in the Sea of
Okhotsk is analyzed from 2005 to 2009. Eddy or non-stationary
component denotes a deflection from its mean value. Because of
strong seasonal signal in the simulated circulation, the monthly
averaging is applied from 2005 to 2009. It is considered four
seasons: winter (January, February, and March), spring (April, May
and June), summer (July, August and September) and autumn (October,
November and December). The EKE is given by the relation
\begin{equation}
\label{eq4} EKE  = \frac{1}{2}\rho_0 \left( \overline{u'^2}+
\overline{v'^2}\right).
\end{equation}
where $\rho_0$ is the density reference, amounting to 1025 kg
m$^{-1}$, and $\overline{u'^2}$, $\overline{v'^2}$ are the monthly
mean squares of the eddy component of zonal and meridional velocity,
respectively. The primes and overbars denote deflections from the
long-term monthly mean and the time averaging, respectively.
$\overline{u'^2}$, $\overline{v'^2}$ are assessed with the relation
\cite{Storch2012}
\begin{equation}
\label{eq5} \overline{x'\cdot y'} = \overline{x\cdot y} -
\overline{x} \cdot
 \overline{y},
\end{equation}
where $x,y$ are the current velocity components (or density
deflection from its reference value), which are extracted from the
model outputs with the time period of 1 day. Besides, kinetic energy
of mean currents (MKE) is assessed with the relation
\begin{equation}
\label{eq6} MKE = \frac{1}{2}\rho_0
\left(\overline{u}^2+\overline{v}^2 \right).
\end{equation}

Fig.~\ref{fig:3} shows the vertical profiles of EKE and MKE are
averaged over the whole basin. In winter, the EKE reaches its
maximal magnitudes, amounting to 15 J m$^{-3}$, near the sea
surface. The EKE magnitudes exceed the MKE magnitudes up to 2.5
times. In summer, EKE and MKE decrease to 8 J m$^{-3}$ and 1.2 J
m$^{-3}$, respectively. Vertical profiles of EKE and MKE indicate
that intense dynamics occurs in the upper 200 m of the Sea of
Okhotsk. Below this layer, EKE and MKE magnitudes reach their low
values, amounting to about 1 J m$^{-3}$, for both seasons. Further,
we will consider EKE variability in the upper 200 m, only.

Fig.~\ref{fig:4} shows the spatial distribution of EKE, integrated
in the upper 200 m, during different seasons. In winter, EKE reach
its maximum value, amounting to 3.5$\times$10$^3$ J m$^{-2}$, along
the western boundary of the Sea of Okhotsk and in the south-western
Kuril Basin. It should be noted that a region with maximal
magnitudes of EKE, ranging from 0.9$\times$10$^3$ to
2.2$\times$10$^3$ J m$^{-2}$, covers a wide area along the western
boundary of the Sea of Okhotsk, which results from different places
of mesoscale eddy generation. A region, situated northward
52$^{\circ}$N, where EKE magnitudes range from 0.45 to
0.9$\times$10$^3$ J m$^{-2}$, widens from 144$^{\circ}$E to
149$^{\circ}$E due to strong hydrodynamic instability of the ESC.
Over the Kuril Basin, along the eastern boundary and on the
north-eastern of the Sea of Okhotsk, EKE magnitudes are limited by
0.45$\times$10$^3$ J m$^{-2}$. In spring, the intensity of mesoscale
variability decreases over the whole basin. Along the western
boundary of the Sea of Okhotsk, EKE decreases up to
0.9$\times$10$^3$ J m$^-{2}$. Decreasing EKE up to
0.75$\times$10$^3$ J m$^{-2}$ is observed in the south-western Sea
of Okhotsk. A small region with the EKE magnitudes, exceeding
1.5$\times$10$^3$ J m$^{-2}$, is situated northward 44$^{\circ}$N.
In other regions of this sea, the EKE magnitudes are limited by
0.3$\times$10$^3$ J m$^{-2}$. In summer, EKE magnitudes shows
minimal values, ranging from 0.15$\times$10$^3$ to 0.3$\times$10$^3$
J m$^{-2}$, for the whole basin. Along the western boundary of the
Sea of Okhotsk, the EKE magnitudes are limited by 0.6$\times$10$^3$
J m$^{-2}$. In autumn, the intensity of mesoscale variability
increases again in the Sea of Okhotsk. The spatial distribution of
EKE shows its maximal magnitudes, amounting to up 2$\times$10$^3$ J
m$^{-2}$. Along the western boundary, in the southern basin and
along the eastern part of this sea, the EKE magnitudes reach to
0.6$\times$10$^3$ J m$^{-2}$.

Thus, the intense mesoscale variability in the Sea of Okhotsk is
observed in upper 200 m, where the EKE magnitudes exceed more than
2.5 times those of MKE in winter and about 9 times in summer. The
spatial distribution of EKE revels the pronounced seasonally varying
EKE with its maximal values in winter and its minimal values in
summer. According to the spatial distributions of EKE, its maximal
values appear along the western boundary of the Sea of Okhotsk,
where the ESC extends, during different seasons. According to the
natural observations \cite{Mizuta2003} and results of previous
numerical simulations \cite{Smizu2006,Matsuda2015}, the transport of
the ESC is characterized by strong seasonal variability. This
transport reaches its maximal values in winter and its minimal
values in the end of summer. It is supposed that seasonal
variability of mesoscale variability can be results from the
hydrodynamic instability of the ESC.

\section{Energy conversion in the Sea of Okhotsk}
\label{sec:4}

According to results of the previous section, the EKE maximal
magnitudes were observed along the western boundary of the Sea of
Okhotsk, where the ESC extends, during different seasons. It is
supposed that hydrodynamics instability (baroclinic and barotropic)
of the along-shore branch of the ESC induces intense mesoscale
variability associated with high EKE magnitudes. To examine this
supposition, two quantities are analyzed in this section. The first
quantity (\textit{BC}) estimates quantitatively the rate of energy
conversion from the mean available potential energy (MPE) to eddy
available potential energy (EPE) and characterizes the baroclinic
instability of the ESC. The second quantity (\textit{BT}) is linked
with the rate of energy conversion from MKE to EKE and characterizes
the barotropic instability of the ESC. To estimate the \textit{BC},
it is used the following relation
\cite{Thomson1984,Eden2002,Zhan2016}

\begin{equation}
\label{eq7} BC = -\frac{g^2}{\overline{N}^2\rho_0}
\overline{\mathbf{u'}_h\rho'}\cdot \nabla_h \overline{\rho},
\end{equation}
where $\nabla_h$ is the horizontal operator, $g$ is the
gravitational acceleration, $\overline{N}^2$ is the basin-averaged
square of the buoyancy frequency \cite{Storch2012}, $\mathbf{u}_h$
is the horizontal velocities and $\rho$ is the density deflection
from the reference value $\rho_0$. To assess an eddy density flux
$(\overline{\mathbf{u'}_h\rho'})$, it was used the relation
(\ref{eq5}). According to (\ref{eq7}), negative \textit{BC}
indicates that EPE is converted to MPE, when
$\overline{\mathbf{u'}_h\rho'}$ is directed in the same direction
with the horizontal gradient of mean density. On the other hand,
when the \textit{BC}$>$0, then $\overline{\mathbf{u'}_h\rho'}$ is
against the direction of mean density gradient, that is, MPE is
converted to EPE.

Fig.~\ref{fig:5} shows spatial distributions of the \textit{BC},
integrated in the upper 200 m during different seasons. According to
these distributions, maximal magnitudes of the \textit{BC} occur
along the western boundary of the Sea of Okhotsk in winter, when the
rate of energy conversion from MPE to EPE exceeds to
6$\times$10$^{-2}$ W m$^{-2}$. In the other regions of the Sea of
Okhotsk, the \textit{BC} magnitudes are two times less than those
along the western boundary of this sea. In spring, the path of
energy conversion from EPE to MPE predominates along the eastern and
western boundaries of the Sea of Okhotsk. The \textit{BC} reaches
its minimal values in summer, when the rate of energy conversion
from MPE to EPE is limited by 10$^{-3}$ W m$^{-2}$. In autumn, high
magnitudes of the \textit{BC}, amounting to 5$\times$10$^{-2}$ W
m$^{-2}$, occur along the western boundary of the Sea of Okhotsk.
However, the autumn distribution of the \textit{BC} is very
heterogeneous and the spots of positive values of the \textit{BC}
alternate with those of negative values of the \textit{BC}. This
heterogeneity of the \textit{BC} distribution does not allow single
out the predominant path of energy conversion with the exception of
a small region northern 52$^{\circ}$N along the western boundary of
the Sea of Okhotsk, where the path of energy conversion from MPE to
EPE predominates.

It should be noted that positive magnitudes of the \textit{BC}
predominate during different seasons in the south-western Kuril
Basin. Despite on low positive magnitudes of the \textit{BC} over
the Kuril Basin, limited by the value of 1-2$\times$10$^{-2}$ W
m$^{-2}$, it indicates that the path of energy conversion from MPE
to EPE predominates. In addition, it coincides with the result of
the previous work \cite{Ohshima2005}, where the predominant
influence of baroclinic instability on mesoscale variability has
been revealed over the Kuril Basin. However, low values of the
\textit{BC} are probably induced by the underestimate of the tidal
mixing, generating the frontal zone in the southern Kuril Basin.

To analyze the contribution of the horizontal shear of the ESC,
associated with barotropic instability, the rate of energy
conversion from MKE to EKE is estimated as following the relation
\cite{Thomson1984,Eden2002,Zhan2016}

\begin{equation}
\label{eq8} BT = - \rho_0 \overline{
\mathbf{u'}_h\cdot\left(\mathbf{u'}\cdot
\nabla_h\overline{\mathbf{u}_h} \right) }.
\end{equation}
According to (\ref{eq8}), positive \textit{BT} characterizes the
rate of energy conversion from MKE to EKE and negative \textit{BT}
characterizes the rate of energy conversion from EKE to MKE.

Fig.~\ref{fig:6} shows the spatial distribution of the \textit{BT},
integrated in the upper 200 m during different seasons. According to
these distributions, maximal magnitudes of the \textit{BT},
amounting to about 3$\times$10$^{-3}$ W m$^{-2}$ and
2$\times$10$^{-3}$ W m$^{-2}$, are observed  in winter and in
autumn. Minimal absolute magnitudes of the \textit{BT} are limited
by 5$\times$10$^{-4}$ W m$^{-2}$ in summer. The maximal magnitudes
of the \textit{BT} are observed in the north-western part of the Sea
of Okhotsk, where isobaths are thickening and high horizontal shear
of the ESC occurs. This distribution of the \textit{BT} is spatial
heterogeneity, characterized by alternation of the regions with
energy conversion from MKE to EKE and the regions with the energy
conversion from EKE to MKE.

Comparing spatial distributions of the \textit{BC} (see,
fig.~\ref{fig:5}) and \textit{BT} (see, fig.~\ref{fig:6}) indicates
that from winter to spring along the western boundary of the Sea of
Okhotsk the energy conversion from MPE to EPE predominates over the
energy conversion from MKE to EKE. Spatial distributions of the
\textit{BC} are more uniform in contrast to those of the
\textit{BC}. Heterogeneity of the \textit{BT} distribution reduces
its integral contribution to EKE in contrast to the integral
contribution of the \textit{BC}. Note that the region with maximal
magnitudes of the \textit{BT} is narrower than the region with
maximal magnitudes of the \textit{BC} along the western boundary of
the Sea of Okhotsk. Thus, the presented results indicate that the
barotropic instability and baroclinic instability of the along-shore
branch of the ESC can be responsible for mesoscale variability along
the western boundary of the Sea of Okhotsk.

\section{Eddy energy budget in the Sea of Okhotsk}
\label{sec:5}

At examining EKE in the closed basins, a general framework is based
on an analysis of the EKE budget equation as proposed by
\cite{Storch2012}. At considering the terms of the EKE budget
equation, we can assess sinks and sources of the EKE and its
dissipation as well as the energy conversions between various
components of the total energy. These estimates are very important
at analyzing heat and fresh water budgets as well as forecasting the
ecosystem evolution in the Sea of Okhotsk. In this study, impacts of
hydrodynamic instability of the ESC and wind power input are
considered as the major sources of the EKE in the Sea of Okhotsk.

The system equations for ocean circulation, formulated in the
Boussinesq and hydrostatic approximations, has a form

\begin{equation}
\label{eq9} \left\{ \begin{array}{l}
\frac{{d{{\bf{u}}_h}}}{{dt}} + f{\bf{k}} \times {{\bf{u}}_h} + \frac{{{\nabla _h}p}}{{{\rho _0}}} = \frac{{{{\mathbf{F}}_h}}}{{{\rho _0}}}\\
\frac{{\partial p}}{{\partial z}} =  - \rho g\\
{\nabla _h} \cdot {\mathbf{u}_h} + \frac{{\partial w}}{{\partial z}}
= 0
\end{array} \right.
\end{equation}
Here $w$ is the vertical velocity, $\mathbf{k}$ is the vertical
single vector, $p$ is the pressure and $\mathbf{F}_h$ is the
external forcing.

According to \cite{Zhai2013}, the solution of equation (\ref{eq9})
can be presented as a sum of two components: time-mean and
time-varying (turbulent) components. The EKE budget equation
(\ref{eq4}) with its sources and sinks as well as energy conversion
paths has a form
\begin{equation}
\label{eq10}\nabla \cdot \overline {p'{\mathbf{u'}}}  + \frac{{{\rho
_0}}}{2}\nabla  \cdot \overline {\left( {{\mathbf{u}} \cdot
{\mathbf{u'}^2_h}} \right)}  + {\rho _0}\overline {\frac{\partial
}{{\partial t}}\frac{\mathbf{u'}^2_h}{2}}  =  - \overline {\rho 'w'}
g + \overline {{{{\mathbf{u'}}}_h} \cdot \mathbf{F'}_h} - {\rho
_0}\overline {{{{\mathbf{u'}}}_h} \cdot \left( {{\mathbf{u'}} \cdot
\nabla \overline {{{\mathbf{u}}_h}} } \right)}
\end{equation}
where $\mathbf{u}$ is the three-dimensional velocity field.

In equation (\ref{eq10}), the first term on the right-hand side
(RHS), $-\overline{\rho'w'}g$, denotes the rate of energy conversion
from EPE to EKE and measures the strength of baroclinic instability.
The second term on the RHS, $\overline{\mathbf{u'}_h \cdot
\mathbf{F'}_h}$ denotes the time-varying component of wind forcing
and internal turbulent viscosity induced by the sub-grid processes.
Last term on the RHS, $-\rho_0\overline{\mathbf{u'}_h \cdot
(\mathbf{u}' \cdot \nabla \overline{\mathbf{u}_h})}$, denotes the
kinetic energy exchange between the mean current and eddies and
corresponds to \textit{BT}$+\left(-\overline{u'w'}\frac{\partial
\overline{ u}}{\partial z} -\overline{v'w'}\frac{\partial \overline{
v}}{\partial z}\right)$, where the last term presents the
contribution of the vertical shear instability of the mean current.
This term is small in comparison with the \textit{BT}. The first
term of equation (\ref{eq10}) on the left-hand side (LHS),
$-\nabla\cdot\overline{p'\mathbf{u'}}$, denotes the pressure work.
The second term on the LHS, $-\frac{\rho_0}{2}\nabla \cdot
(\overline{\mathbf{u}\cdot\mathbf{u'}_h^2})$, characterizes the
change of the EKE induced by mean current advection; the third term
on the LHS, $-\rho_0\overline{\frac{\partial}{\partial
t}\frac{\mathbf{u'}^2_h}{2}}$, denotes the tendency of the EKE.

\subsection{Sources of the EKE in the Sea of Okhotsk}
\label{sec:5.1}

According to \cite{Smizu2006}, wind stress plays the leading role in
the generation of the basin-scale circulation in the Sea of Okhotsk.
The wind stress curl promotes to generate the cyclonic gyre in the
central part of the Sea of Okhotsk and the alongshore wind stress
component induces the along-shore branch of the ESC. Seasonal
variability of the wind stress being the feature of the monsoon
circulation over the Sea of Okhotsk drives strong seasonal
variability of the circulation in the Sea of Okhotsk, in particular,
seasonal variability of the ESC transport \cite{Smizu2006}. It is
supposed that the wind power input can be one of the major sources
of the EKE and mesoscale variability in the Sea of Okhotsk.

According to the EKE budget equation (\ref{eq10}), the one of the
sources of the EKE, ($\overline{\mathbf{u'}_h\cdot\mathbf{F'}_h}$)
is the wind power input. Wind power input can be assessed by
following relation \cite{Huang2006,Zhai2012,Wunsch1998}.

\begin{equation}
\label{eq11} G  = \overline {{\tau _x}\cdot{u_s}}  + \overline
{{\tau _y}\cdot{v_s}},
\end{equation}
where $\tau_x$ ,$\tau_y$ are the wind stress components and $u_s$,
$v_s$ are the zonal and meridional velocities on the sea surface,
respectively. The relation (\ref{eq11}) can be presented as
\begin{equation}
\label{eq12}
\begin{array}{l}
G  = \overline {{\tau_x}\cdot{u_s}}  + \overline {{\tau _y}\cdot{v_s}}  = {G_1} + {G_2},\\
{G_1} = \overline {{\tau _x}}  \cdot \overline {{u_s}}  + \overline
{{\tau _y}}  \cdot \overline {{v_s}} ,\,\,{G_2} = \overline {{{\tau
'}_x}\cdot{{u'}_s}}  + \overline {{{\tau '}_y}\cdot{{v'}_s}}.
\end{array}
\end{equation}
Here, $G_1$ denotes the rate of energy conversion from wind energy
to MKE and $G_2$ denotes the rate of energy conversion from wind
energy to EKE, where $\tau'_x,\tau'_y$ denote a time-varying
component of wind stress. Positive $G_2$ indicates that the
time-varying component of wind stress promotes to increase EKE on
the sea surface and negative $G_2$ indicates that the time-varying
component of wind stress prevents to increase EKE on the sea
surface. It should be noted that in contrast to the approach,
presented in the works \cite{Wunsch1998,Zhai2012,Huang2006}, instead
geostrophic velocity, the sea surface velocities are used.

According to the spatial distribution of $G_2$, it reaches its
maximal magnitudes, amounting to 4$\times$10$^{-2}$ W m$^{-2}$ in
winter and 3$\times$10$^{-2}$ W m$^{-2}$ in autumn (see,
fig.~\ref{fig:7}). In winter, intensive energy exchange between the
time-varying components of wind stress and EKE occurs in the
north-eastern and western Sea of Okhotsk. In spring, the $G_2$
magnitudes decrease significantly from 4$\times$10$^{-2}$ W m$^{-2}$
to 0.6$\times$10$^{-2}$ W m$^{-2}$. However, high magnitudes of
$G_2$, amounting to 0.9-1.2$\times$10$^{-2}$ W m$^{-2}$, occur along
the western and eastern boundaries of the Sea of Okhotsk. In summer,
the $G_2$ magnitudes decrease and reach their minimal values,
amounting to 0.3$\times$10$^{-2}$ W m$^{-2}$, over the whole basin,
except a small region, situated along the western boundary of this
basin, where the $G_2$ magnitudes amount to 1.2$\times$10$^{-2}$ W
m$^{-2}$. In autumn, the intensity of energy exchange between the
time-varying wind stress component and EKE increases up to
2.5$\times$10$^{-2}$ W m$^{-2}$ in the northern, western and eastern
parts of the Sea of Okhotsk.

Thus, $G_2$ exhibits high positive values along the western boundary
of the Sea of Okhotsk, amounting to 4$\times$10$^{-2}$ W m$^{-2}$ in
winter and 1.2$\times$10$^{-2}$W m$^{-2}$ in summer, that is, the
time-varying wind stress promotes to increase the EKE. High values
of the $G_2$ in comparison with the \textit{BC} values (see,
fig.~\ref{fig:5}) and the \textit{BT} values (see, fig.~\ref{fig:6})
indicates the leading role of the time-varying wind stress in the
generation of the EKE along the western boundary of the Sea of
Okhotsk.

As following from the previous section~\ref{sec:4}, the \textit{BC}
magnitudes exceed those of the \textit{BT}. According to
\cite{Zhan2016}, the mechanism of baroclinic instability consists of
two stage. On the first stage, it is realized energy conversion from
MPE to EPE. On the second stage, EPE converts to EKE. The intensity
of this energy conversion is characterized by the magnitude of the
source of the EKE budget equation (\ref{eq10}), which is given by
the relation
\begin{equation}
\label{eq13} -\overline{\rho'w'}g,
\end{equation}
where $w'$ is the time-varying vertical velocity and $\rho'$ is the
time-varying density component. Positive values of
$-\overline{\rho'w'}g$ point out denser (later) water masses
associated with downward (upward) movements.

Fig.~\ref{fig:8} shows a spatial distribution of
$-\overline{\rho'w'}g$, integrated in the upper 200 m during
different seasons. According to this distribution,
$-\overline{\rho'w'}g$ magnitudes reach its maximal values,
amounting to 6$\times$10$^{-3}$ W m$^{-2}$, along the western
boundary of the Sea of Okhotsk in winter. The region with the
$-\overline{\rho'w'}g$ maximal magnitudes covers the shelf zone on
the western boundary of this sea. In spring, the rate of energy
conversion from EPE to EKE decreases up to 3$\times$10$^{-3}$ W
m$^{-2}$ in the south-western part of the Sea of Okhotsk; the
spatial distribution of $-\overline{\rho'w'}g$ is strongly
heterogeneous. In summer, the $-\overline{\rho'w'}g$ minimal
magnitudes, amounting to 10$^{-3}$ W m$^{-2}$, are observed over the
whole basin. In autumn, the rate of energy conversion from EPE to
EKE increases again along the western boundary of the Sea of Okhotsk
and reaches its winter-time magnitudes. However, the spatial
distribution of $-\overline{\rho'w'}g$ is strongly heterogeneous in
contrast to that in winter. In addition, this spatial distribution
covers less width shelf region than that in winter.

Thus, our analysis of the EKE sources shows that the time-varying
wind stress component predominates on the other sources of the EKE
in the Sea of Okhotsk. However, the pronounced energy conversion
from MPE to EPE in comparison with the energy conversion from MKE to
EKE in winter and in autumn, as well as high magnitudes of
$-\overline{\rho'w'}g$, indicate importance of baroclinic
instability in the generation of mesoscale variability along the
western boundary of the Sea of Okhotsk.

\section{Hydrological conditions and mesoscale eddies on the eastern shelf of Sakhalin Island from winter to spring 2005}
\label{sec:6}

In the previous sections, it was showed that the ESC exhibits
hydrodynamic instability in winter. In this section, it is presented
the results of the baroclinic instability of the along-shore branch
of the ESC from January to May 2005 on the eastern shelf of Sakhalin
Island (see, fig.~\ref{fig:1}). To demonstrate the results of the
baroclinic instability of the ESC, hydrodynamic conditions are
considered along the western boundary of the Sea of Okhotsk.

At first, we consider the meridional velocity ($v$) and deviation of
density ($\rho$) from the reference value along the western boundary
of the Sea of Okhotsk. Fig.~\ref{fig:9} shows vertical sections of
the monthly mean of $v$ and $\rho$ across the shelf on
50.46$^{\circ}$N from January to March 2005.

As following from the numerical simulations, the narrow region
characterized by $v$ magnitudes, ranging from 0.16 to 0.4 m
s$^{-1}$, is observed in January 2005 (see, fig.~\ref{fig:9}(a)).
From February to March, this region extends to offshore and deepens
from 30 to 60 m (see, fig.~\ref{fig:9}(b)--(c)). At the same time,
the $\rho$ vertical section shows outcropping isopycnals (see,
fig.~\ref{fig:9}(a)--(c)). These features of hydrodynamic conditions
are observed along the western boundary of the Sea of Okhotsk.

Let us to give evidence that the presented features of vertical
change of $v$ relate to deformation of isopycnal surfaces. According
to \cite{Pedlosky1987} in the geostrophic and hydrostatic
approximations, the vertical shear of $v$ is balanced by the zonal
density gradient along isobaths
\begin{equation}
\label{eq15} \frac{\partial v}{\partial z}  = - \frac{g}{\rho_0 f_0
R \cos \vartheta} \left( \frac{\partial \rho}{\partial
\theta}\right)_p.
\end{equation}
Here $f_0=2\Omega\sin\vartheta_0$ is the Coriolis parameter at the
given latitude $\vartheta_0$, $\theta$ is the longitude and $R$ is
the Earth radius.

Let us to estimate the relation (\ref{eq15}) along the western
boundary of the Sea of Okhotsk from January to March 2005. In
January, the $\partial v/\partial z$ maximal magnitudes are observed
in a narrow region along this boundary, where maximal magnitudes of
the zonal density gradient occur (see, fig.~\ref{fig:9}(d)). From
February to March, the area of this region increases. An analysis of
relation (\ref{eq15}) on other vertical sections (not shown) across
the eastern shelf of Sakhalin Island shows that the right-hand side
(\textit{RHS}) and the left-hand side (\textit{LHS}) of this
relation are very similar. Thus, the state of the fluid on the
eastern shelf of Sakhalin Island is baroclinic from January to
March.

Let us to consider the velocity field along the western boundary of
the Sea of Okhotsk. It is found that eddy-like structures are
generated in the velocity field from February to May 2005. In the
field of the vertical component of relative vorticity vector
(relative vorticity) $\left(\omega= (\partial v/\partial\theta
-\partial(u\cos\vartheta)/\partial\vartheta)/(Rf \cos\vartheta)
\right)$ these structures are associated with the spots of negative
values of $\omega$. Fig.~\ref{fig:10} shows the velocity and
$\omega$ fields at the horizon of 20 m on 8 April 2005.

According to the presented velocity field, three eddy structures are
observed along the eastern shelf of Sakhalin Island. In the moment
of their generation, the along-shore branch of the ESC follows along
isobaths, ranging from 200 to 240 m, with mean current velocity
amounts to 0.25-0.3 m s$^{-1}$. In the relative vorticity field,
spots with negative magnitudes of the $\omega$, amounting to about
-0.3 and associating with the presented eddy structures, are
observed. Negative magnitudes of the $\omega$ indicate that these
eddy-like structures are anticyclonic eddies.

Fig.~\ref{fig:11} shows vertical structures of the $\omega$, $v$ and
$\rho$ on three zonal sections across the eastern shelf of Sakhalin
Island. The observed eddy structures are characterized by negative
magnitudes of the $\omega$, exhibiting in the upper layer from 50 to
200 m, depending on the zonal section. An analysis of evolution of
these anticyclonic eddy structures shows that they collapse in the
middle of May. Thus, a mean lifetime of these eddies amounts to 45
days. To assess a spatial scale $(L_{eddy})$ of these eddies, a
horizontal scale, where meridional velocity changes its sign, is
estimated. Mean absolute magnitudes of the meridional velocity on
the periphery of these eddies vary from 0.26 to 0.42 m s$^{-1}$.
Hence, the mean spatial scale of these eddy structures varies from
26 to 34 km. This estimation of the $L_{eddy}$ coincides with that
based on the zonal gradient of the $\omega$. Maximal magnitudes of
the $v$ observe from the sea surface to the horizon of 60--80 m.

Let us compare $L_{eddy}$ with $\lambda_1$. To estimate $\lambda_1$,
the boundary value problem (\ref{eq2}) is numerically solved for
monthly mean $N(z)$ profiles, averaged on three zonal sections (see,
fig.~\ref{fig:11}), in April 2005. According to the estimation of
$\lambda_1$, it varies from 8 to 12 km. Thus, $L_{eddy}$ and
$\lambda_1$ are the same order, that is, the observed anticyclonic
eddies are the \textit{mesoscale} anticyclonic eddies.

Thus, along the western boundary of the Sea of Okhotsk during the
winter-spring period, hydrological conditions are characterized by
significant vertical shear of the along-shore velocity, balanced by
the zonal density gradient, and significant horizontal shear of the
ESC with high magnitudes of relative vorticity. From winter to
spring, baroclinic instability of the ESC results in the generation
of the mesoscale eddies on the eastern shelf of Sakhalin Island.

\section{Discussion and Summary}
\label{sec:7}

Based on the outputs of the retrospective numerical simulations in
the north-western Pacific Ocean, the EKE sources were analyzed in
the Sea of Okhotsk. According to our estimation of the first
baroclinic Rossby radius of deformation, the used model resolution
(about 3.5 km) permits mesoscale variability, at least, southward
60$^{\circ}$N. Outputs of these numerical simulations have been
obtained with the INMOM, taking into account the sea ice covering.
Features of mesoscale variability in the Sea of Okhotsk have been
revealed from 2005 to 2009.

Validation of our numerical simulations has showed that the sea
surface velocity field represents main features of the basin-scale
circulation in the Sea of Okhotsk: the East-Sakhalin Current (ESC),
the West--Kamchatka Current, the cyclonic gyre in the central part
of this sea, as well as the anticyclonic circulation over the Kuril
Basin. Comprehensive considering the spatial-temporal structure of
the ESC has showed that this current consists of two branches: the
along-shore branch and the offshore branch. Numerical simulations
have showed the pronounced seasonally varying ESC transport, which
increases up to 6 Sv in winter and decreases up to 1-1.5 Sv in
summer.

The analysis of the EKE, integrated in the upper 200 m, has showed
that the EKE is characterized by the pronounced seasonally varying
with its maximal magnitudes in winter and its minimal magnitudes in
summer. It should be noted that during different seasons the EKE
maximal magnitudes have been observed along the western boundary Sea
of Okhotsk, where the ESC extends. In winter, the EKE magnitudes
increase up to 3.5$\times$10$^3$J m$^{-2}$ along the western
boundary of the Sea of Okhotsk. In spring, the EKE magnitudes vary
from 0.6$\times$10$^3$J m$^{-2}$ to 0.9$\times$10$^3$J m$^{-2}$
along the western boundary of this sea. In summer, the EKE
magnitudes decrease up to 0.3$\times$10$^3$J m$^{-2}$ for the whole
basin. In autumn, the EKE magnitudes increases again up to
2$\times$10$^3$J m$^{-2}$.

As one of the main sources of the EKE, we have considered the
time-varying (turbulent) wind stress component $(\overline{\tau'_x
u'_s}+\overline{\tau'_y v'_s})$. The analysis of the spatial
distribution of $\overline{\tau'_x u'_s}+\overline{\tau'_y v'_s}$
has showed that its maximal magnitudes, amounting to
4$\times$10$^{-2}$ W m$^{-2}$, observe along the western boundary of
the Sea of Okhotsk in winter. In spring and summer,
$\overline{\tau'_x u'_s}+\overline{\tau'_y v'_s}$ magnitudes are
limited by 1.2$\times$10$^{-2}$ W m$^{-2}$. In autumn,
$\overline{\tau'_x u'_s}+\overline{\tau'_y v'_s}$ magnitudes
increase up to 2.5$\times$10$^{-2}$ W m$^{-2}$. Magnitudes of
$\overline{\tau'_x u'_s}+\overline{\tau'_y v'_s}$, integrated over
the Sea of Okhotsk, amount to about 22 GW in winter and about 5.5 GW
in summer. The contribution of the turbulent wind stress exceeds
that of mean wind stress, amounting to about 11 GW in winter and 1
GW in summer.

Because maximal magnitudes of EKE are observed along the western
boundary of the Sea of Okhotsk, we have considered as other sources
of EKE the baroclinic and barotropic instability of the ESC. The
analysis of the rate of energy conversion from MKE to EKE
(\textit{BT}) has showed that \textit{BT} reaches its maximum value,
amounting to 3$\times$10$^{-3}$ W m$^{-2}$, in winter. The
\textit{BT} minimal values, amounting to about 5$\times$10$^{-4}$ W
m$^{-2}$, are observed in the region of the ESC in summer. However,
the distribution of \textit{BT} is strongly heterogeneous and
indicates both on energy conversion from MKE to EKE
(\textit{BT}$>$0) and energy conversion from EKE to MKE
(\textit{BT}$<$0). From spring to summer, the intensity of energy
conversion from MKE to EKE decreases significantly due to the
decrease of the ESC intensity and low horizontal shear of velocity
field in the western Sea of Okhotsk.

To characterize the baroclinic instability of the ESC, we have
considered two variables: \textit{''} and $-\overline{\rho'w'}g$.
The first variable,\textit{BC}, characterizes the intensity of
energy conversion from APE to EPE. The second variable,
$-\overline{\rho'w'}g$, characterizes the intensity of energy
conversion from EPE to EKE. We have established that maximal
intensity of the energy conversion from APE to EPE is observed in
the ESC region in winter. The \textit{BC} magnitudes increase up to
6$\times$10$^{-2}$ W m$^{-2}$. Positive values of \textit{BC}
indicate predominance of energy conversion from APE to EPE in
comparison with energy conversion from EPE to APE. Comparing
\textit{BC} and $\overline{\tau'_x u'_s}+\overline{\tau'_y v'_s}$
points out that these variables are the same order. The intensity of
energy conversion from EPE to EKE $(-\overline{\rho'w'}g)$ reaches
its maximal values, amounting to 6$\times$10$^{-3}$ W m$^{-2}$, in
the western Sea of Okhotsk in winter. Maximal magnitudes of
$-\overline{\rho'w'}g$ cover the whole eastern shelf of Sakhalin
Island. In autumn, the region of the $-\overline{\rho'w'}g$ maximal
values are narrowed.

\begin{figure}
\includegraphics[angle=270,width=1.1\linewidth]{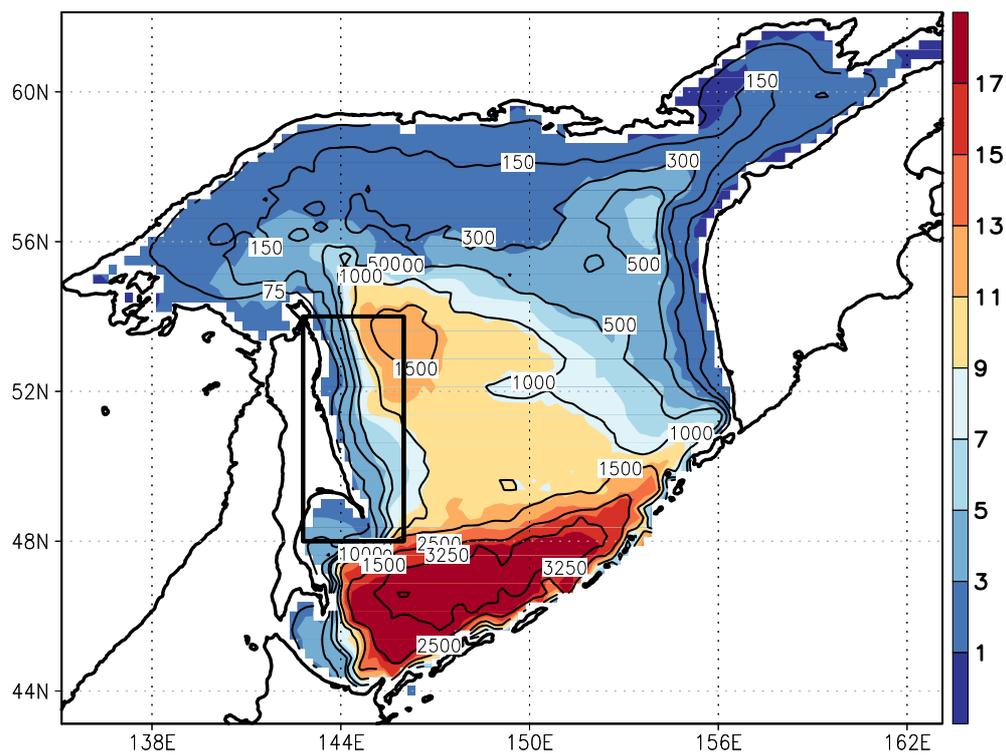}
\caption{Annual mean first baroclinic Rossby radius of deformation
in the Sea of Okhotsk (shading, km). Bottom topography of the Sea of
Okhotsk extracted from the GEBCO dataset and smoothed by 9-point
filter (lines, m). The rectangle marks out the region on the eastern
Sakhalin Island (141.6$^{\circ}$E--146$^{\circ}$E,
44$^{\circ}$N--55$^{\circ}$N).} \label{fig:1}
\end{figure}

\begin{figure}
\includegraphics[angle=270,width=1.4\linewidth]{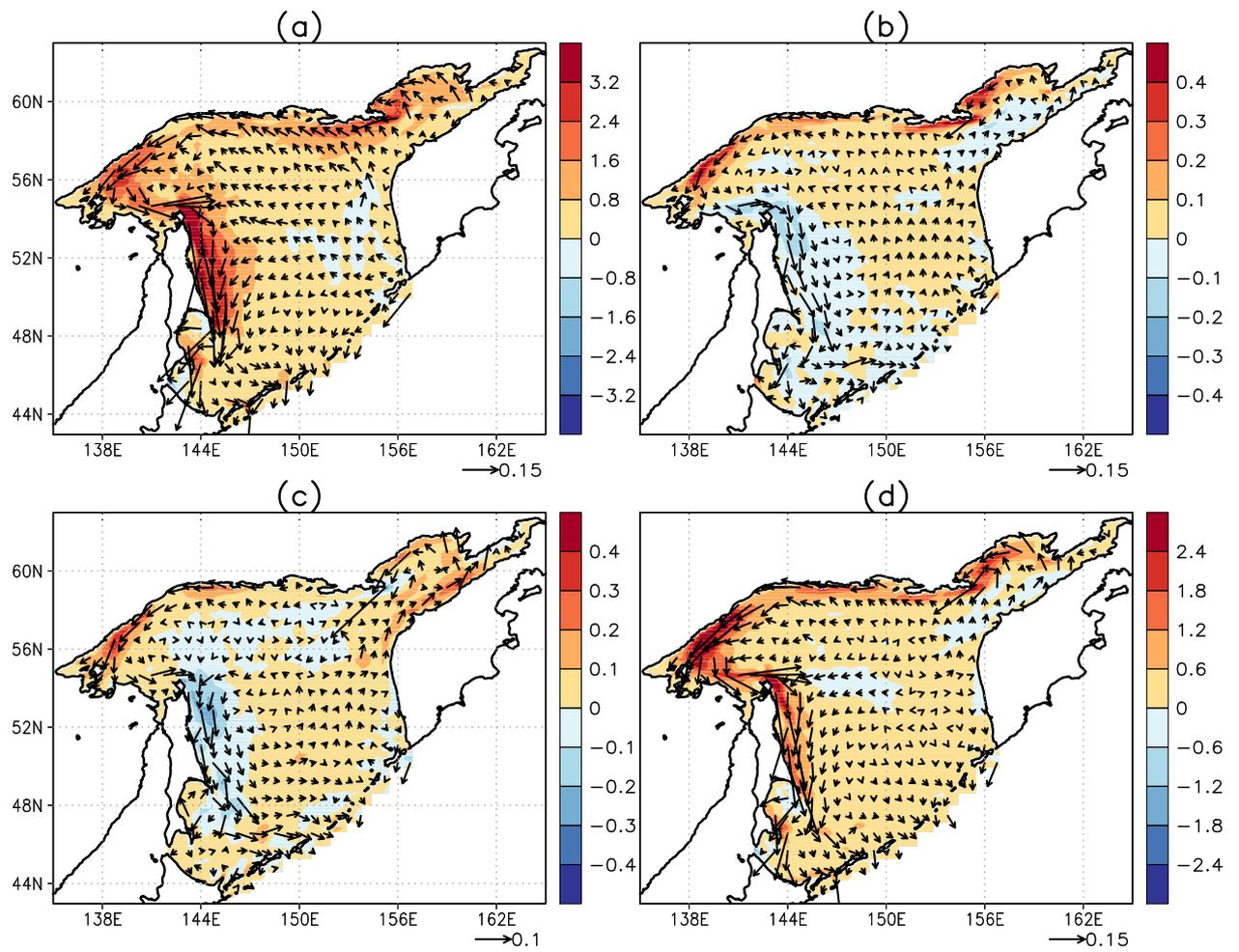}
\caption{Seasonal mean wind power input (shading, 10$^{-2}$W
m$^{-2}$) and velocity field at the horizon of 10 m (vectors, m
s$^{-1}$) in (a) winter (January, February, March), (b) spring
(April, May, June), (c) summer (July, August, September) and (d)
autumn (October, November, December).} \label{fig:2}
\end{figure}

\begin{figure}
\includegraphics[angle=270,width=0.9\linewidth]{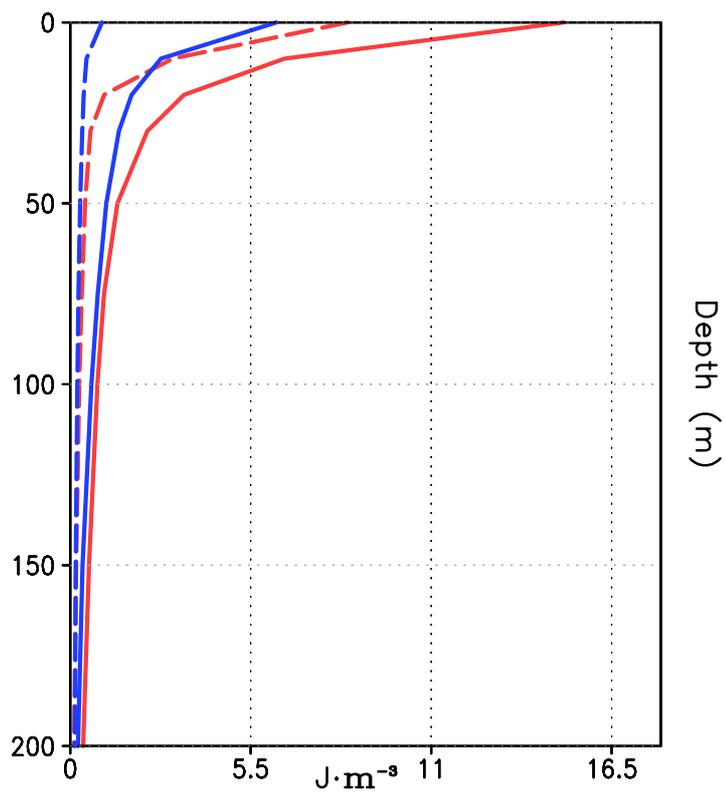}
\caption{Basin-averaged vertical profiles of the EKE (red line) and
MKE (blue line) over the Sea of Okhotsk in winter (January,
February, and March) (dashed line) and summer (July, August and
September) (solid line).} \label{fig:3}
\end{figure}

\begin{figure}
\includegraphics[angle=270,width=1.4\linewidth]{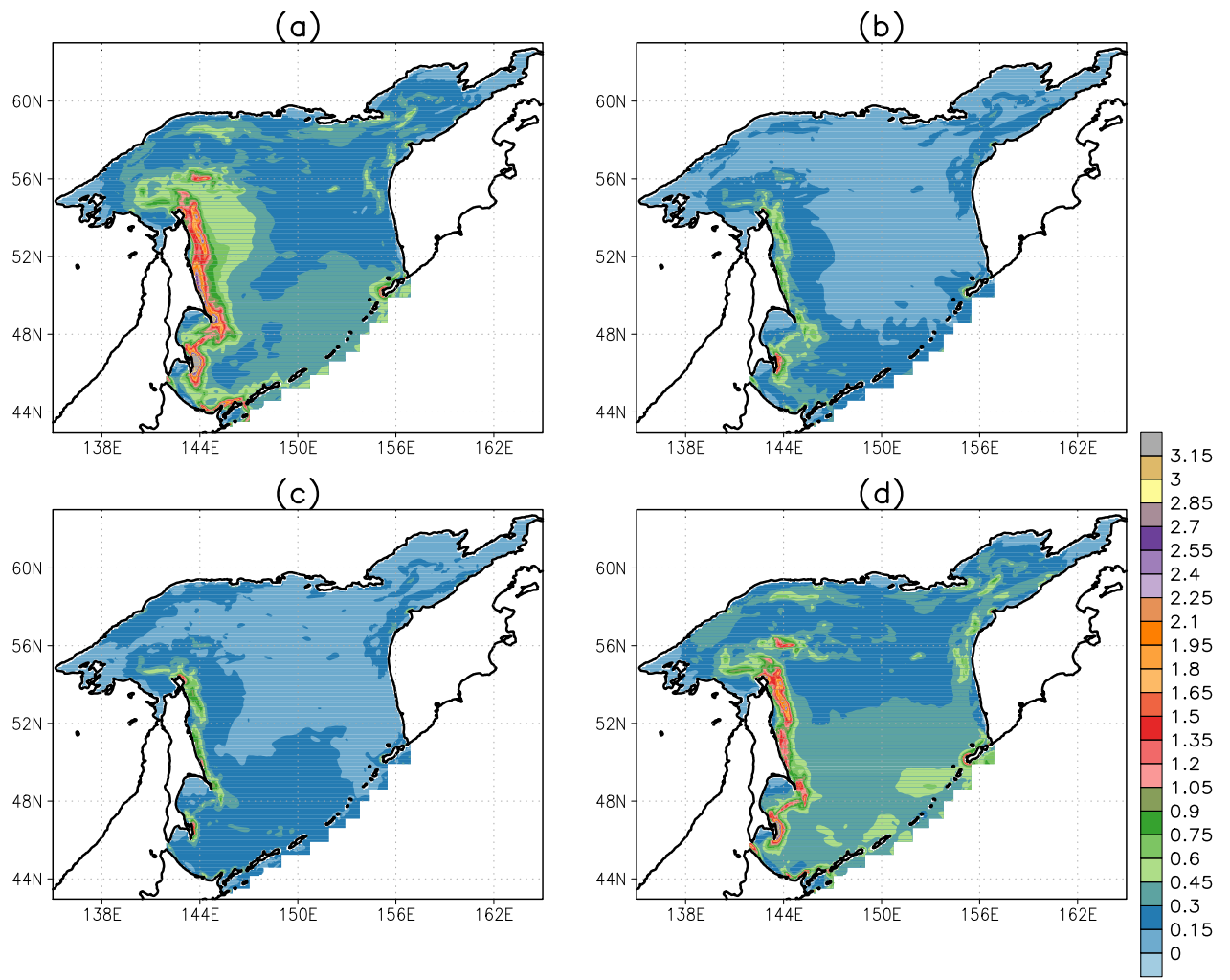}
\caption{The mean EKE (10$^{3}$ J m$^{-2}$) integrated in the upper
200 m in (a) winter, (b) spring, (c) summer and (d) autumn.}
\label{fig:4}
\end{figure}

\begin{figure}
\includegraphics[angle=270,width=1.4\linewidth]{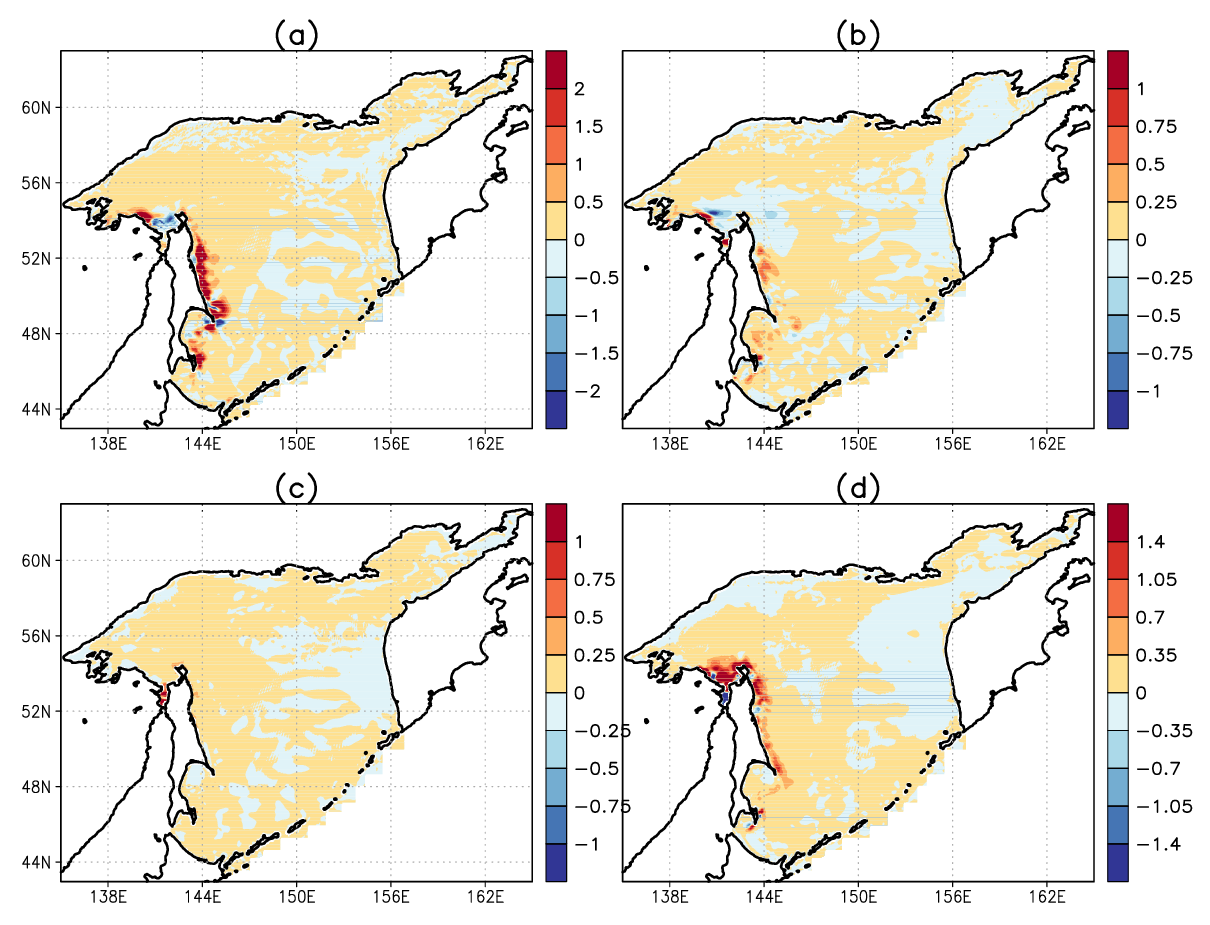}
\caption{Distribution of the rate of the energy conversion term
(\textit{BC}) (10$^{-2}$ W m$^{-2}$) integrated in the upper 200 m
in (a) winter, (b) spring, (c) summer and (d) autumn.} \label{fig:5}
\end{figure}

Tab.~\ref{tab:1} shows magnitudes of \textit{BT}, \textit{BC},
$\overline{\tau'_x u'_s}+\overline{\tau'_y v'_s}$ and
$-\overline{\rho'w'}g$, integrated in the upper 200 m on the eastern
shelf of Sakhalin Island (see, fig.~\ref{fig:1}) during different
seasons.

\begin{figure}
\includegraphics[angle=270,width=1.4\linewidth]{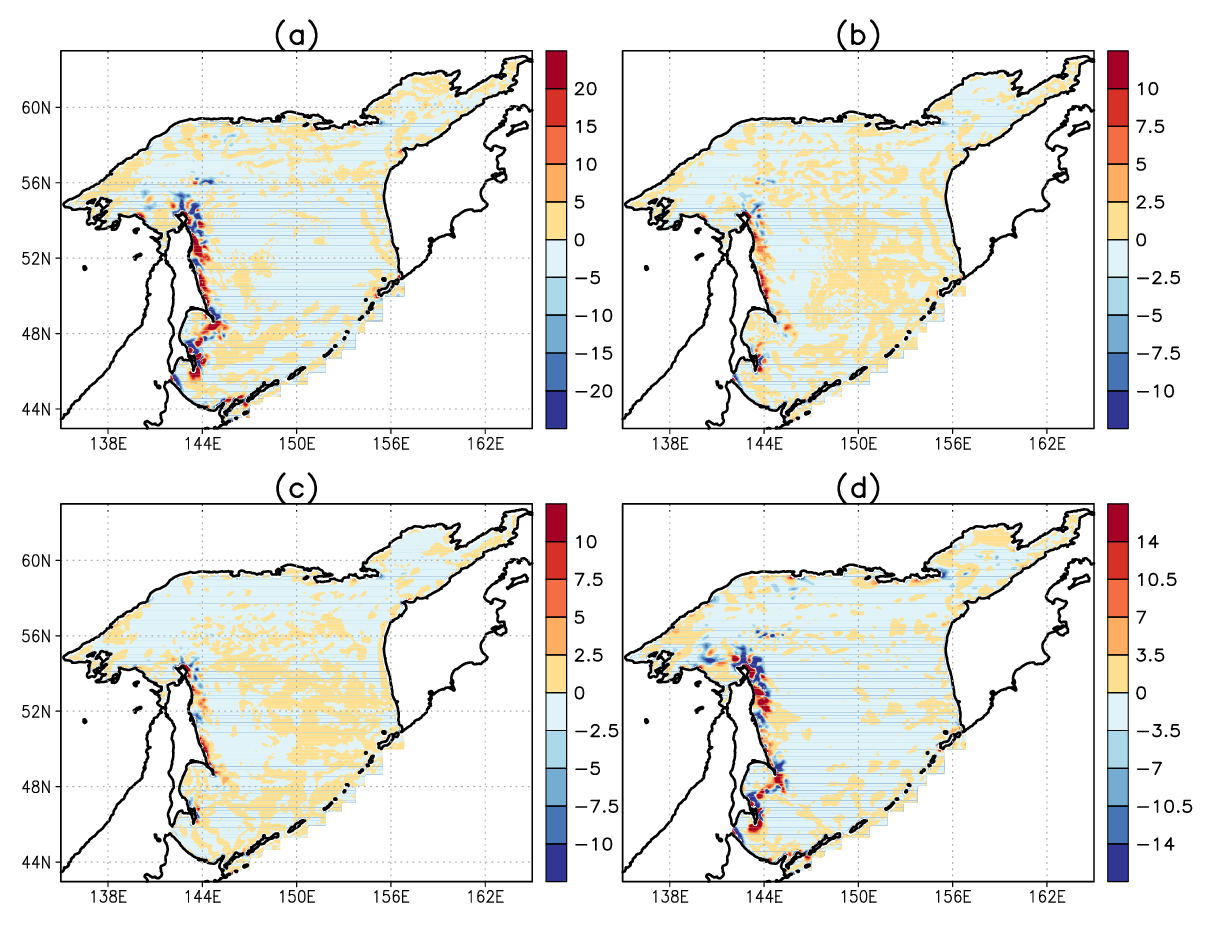}
\caption{Distribution of the rate of the energy conversion term
(\textit{BT}) (10$^{-4}$ W m$^{-2}$) integrated in the upper 200 m
in (a) winter, (b) spring, (c) summer and (d) autumn.} \label{fig:6}
\end{figure}

\begin{figure}
\includegraphics[angle=270,width=1.4\linewidth]{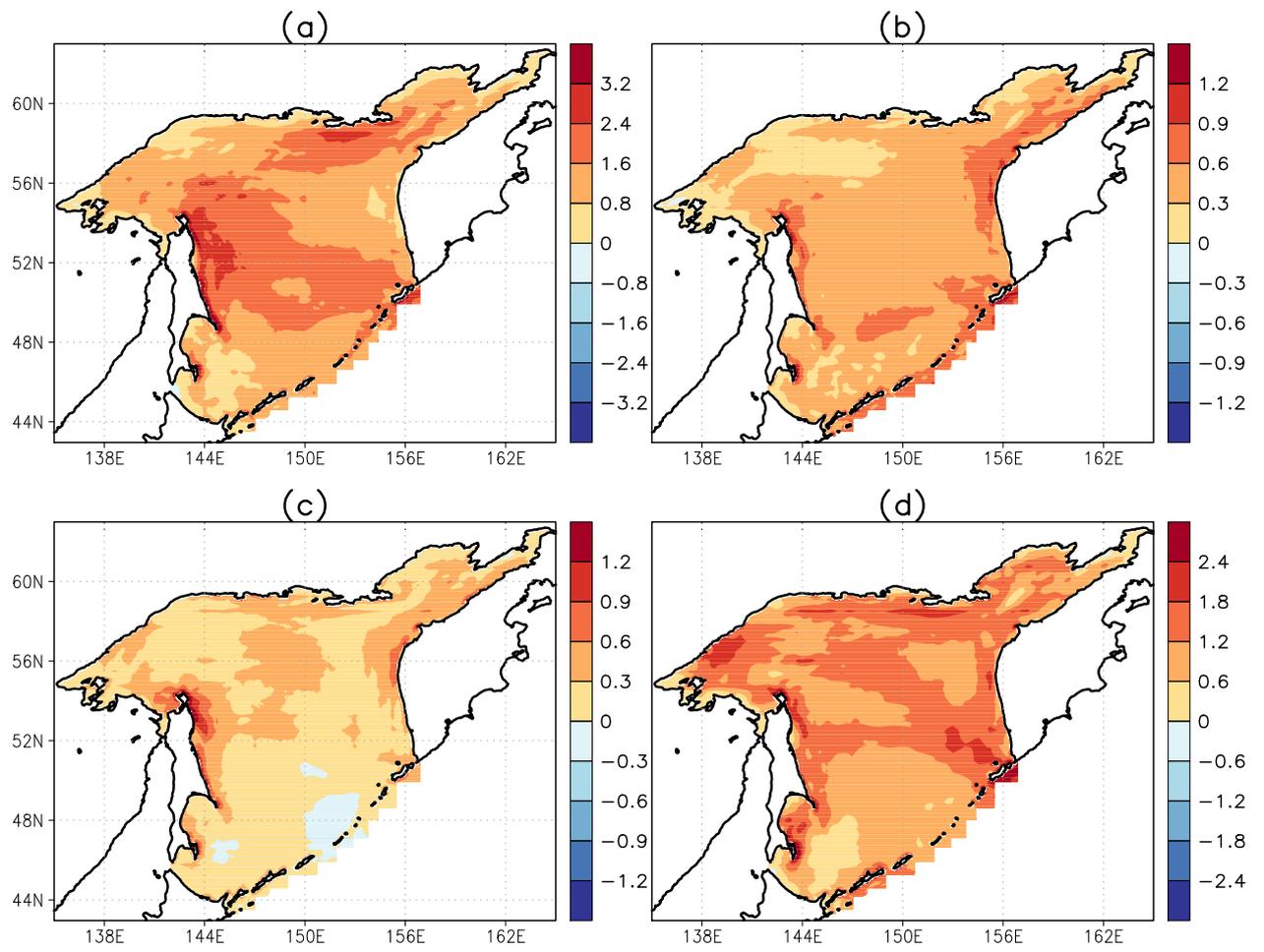}
\caption{Distribution of generation of EKE due to time-varying wind
stress (10$^{-2}$ W m$^{-2}$) in (a) winter, (b) spring, (c) summer
and (d) autumn.} \label{fig:7}
\end{figure}

\begin{figure}
\includegraphics[angle=270,width=1.4\linewidth]{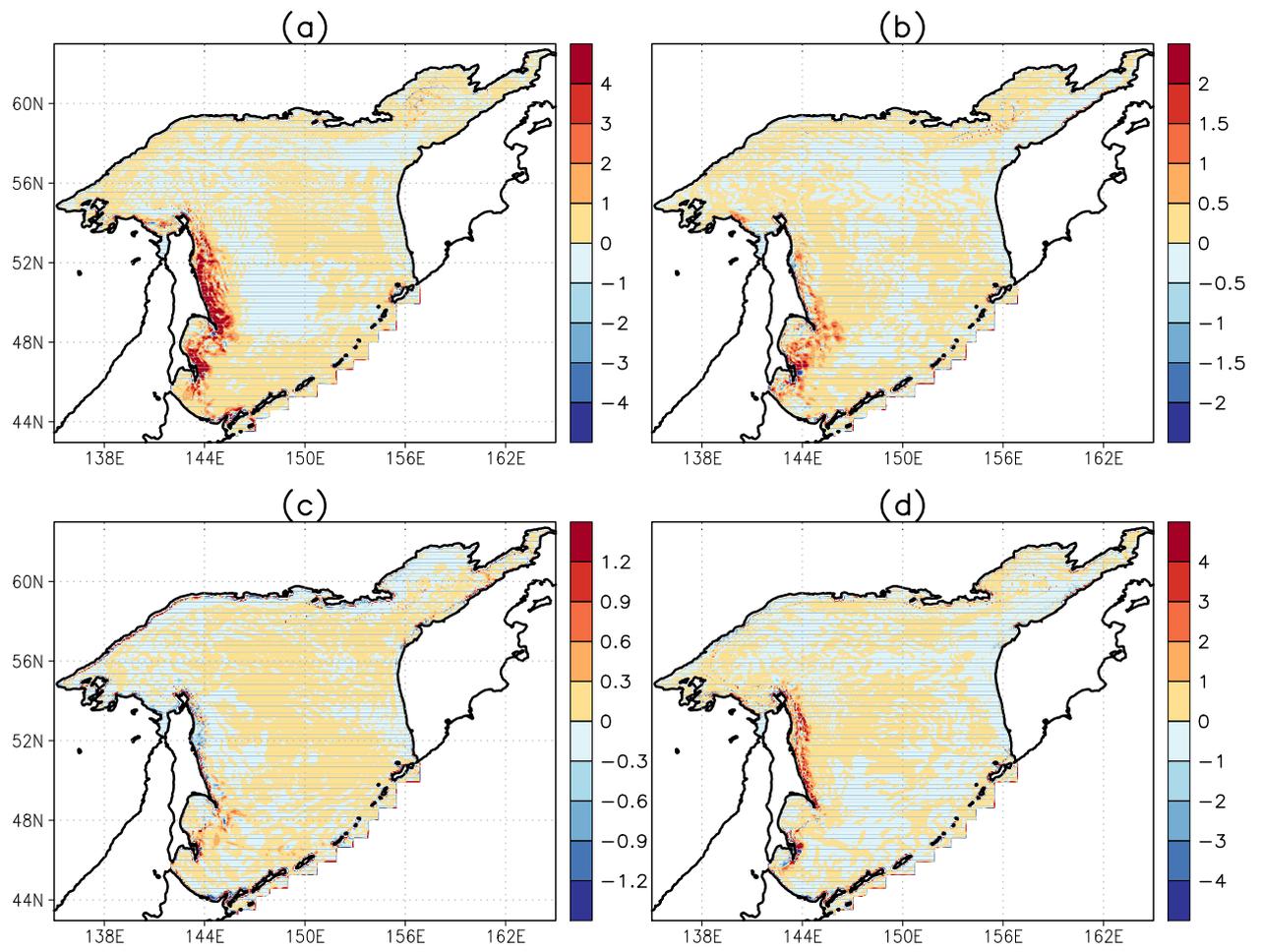}
\caption{Distribution of the rate of energy conversion from EPE to
EKE $(-\overline{\rho'w'}g)$ (10$^{-3}$ W m$^{-2}$) integrated in
the upper 200 m in (a) winter, (b) spring, (c) summer and (d)
autumn. } \label{fig:8}
\end{figure}

\begin{figure}
\includegraphics[angle=270,width=1.4\linewidth]{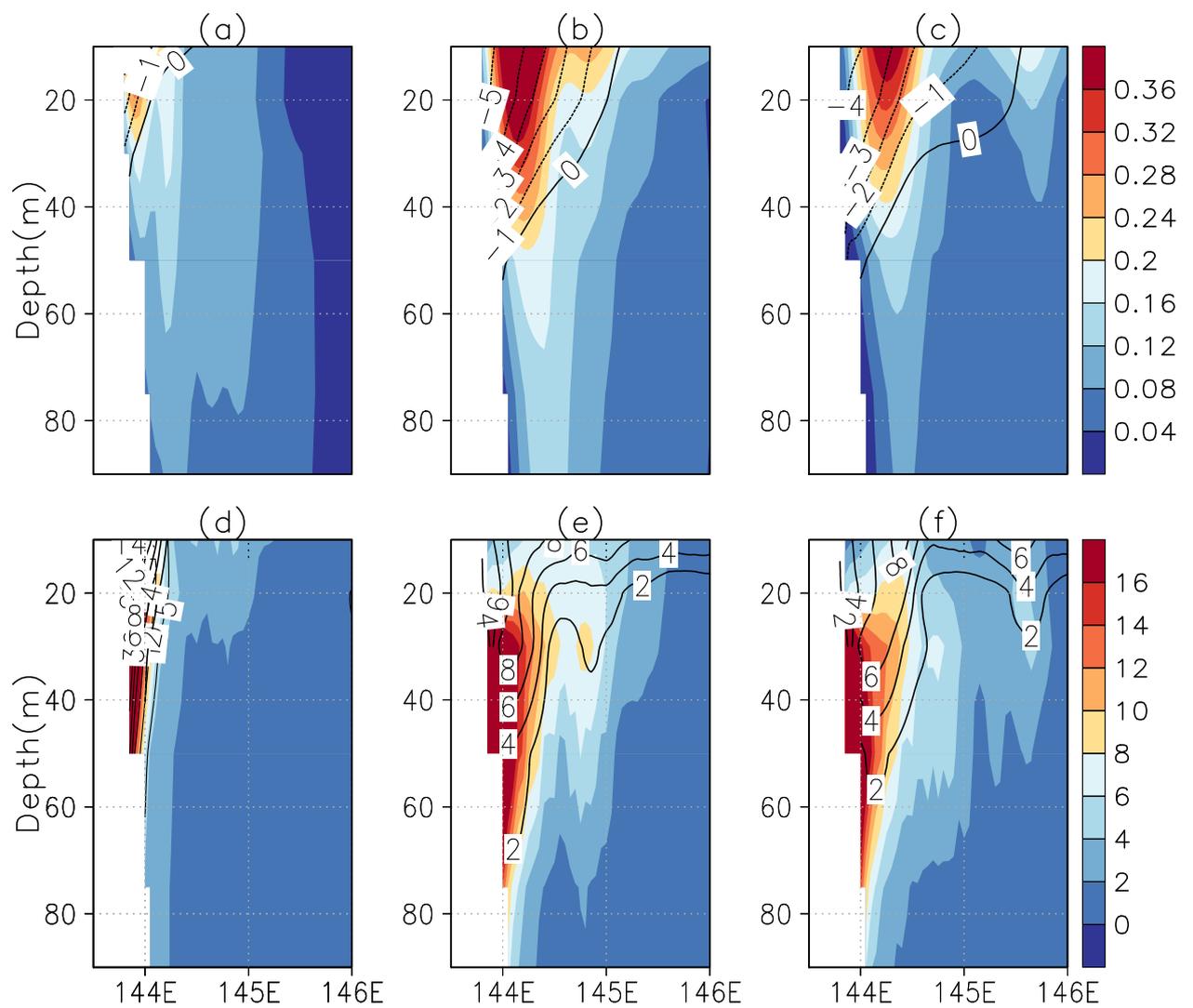}
\caption{Vertical section of monthly mean meridional velocity
(shedding, m s$^{-1}$) and density deviation (line, kg m$^{-3}$)
from the reference value $\rho_0$, amounting to 1025 kg m$^{-3}$,
across the eastern shelf of Sakhalin Island (50.46$^{\circ}$N) in
(a) January, (b) February and (c) March 2005. Left-hand side (lines,
10$^{-3}$ s$^{-1}$) and right-hand side (shading, 10$^{-3}$
s$^{-1}$) of relation (15) in: (d) January, (e) February, and (f)
March 2005.} \label{fig:9}
\end{figure}

\begin{figure}
\includegraphics[width=0.8\linewidth]{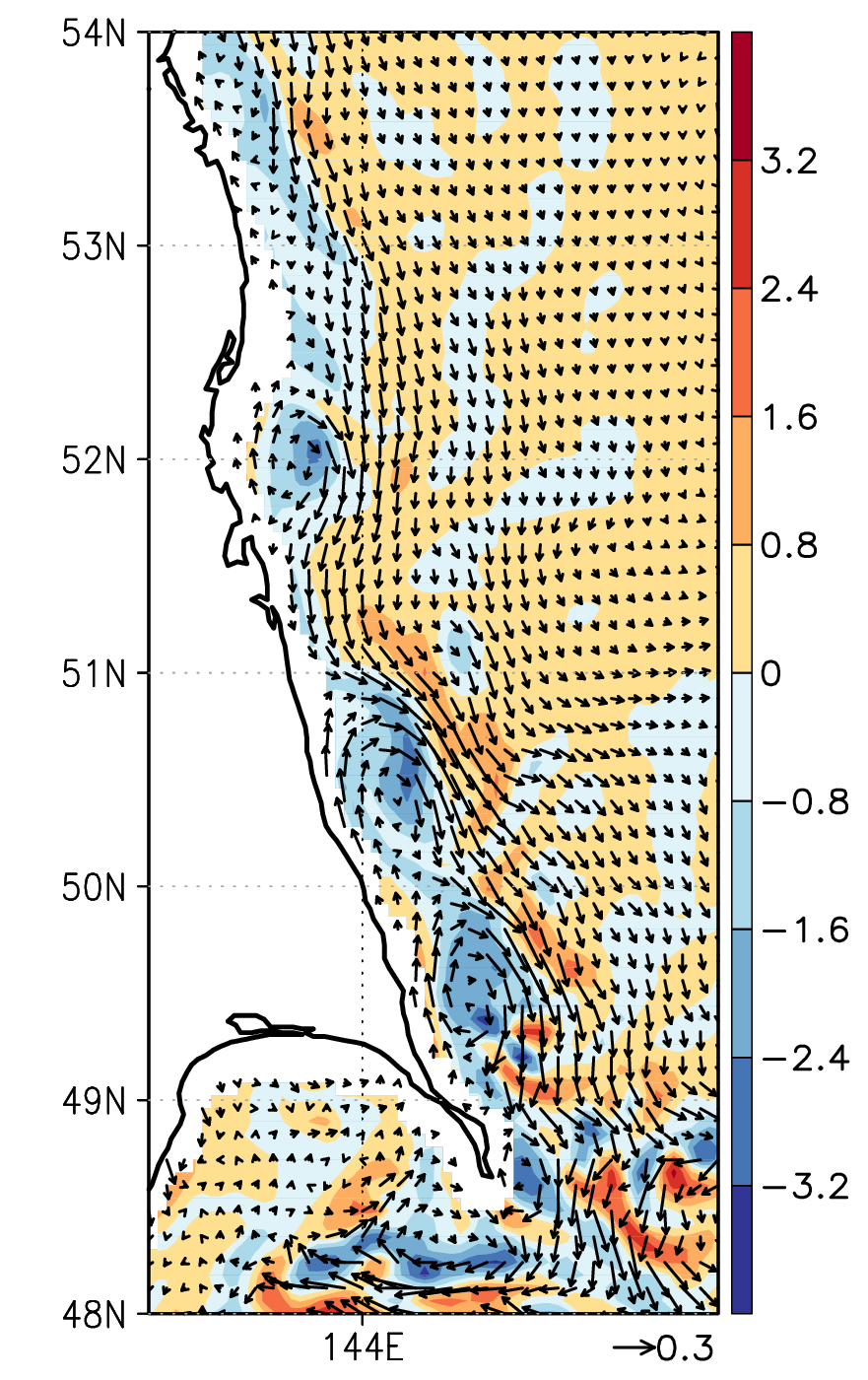}
\caption{Velocity field (vectors, m s$^{-1}$) and vertical component
of relative vorticity (shading, 10$^{-1}$) at the horizon of 20 m on
the eastern shelf of Sakhalin Island on 8 April 2005.}
\label{fig:10}
\end{figure}

\begin{figure}
\includegraphics[angle=270,width=1.4\linewidth]{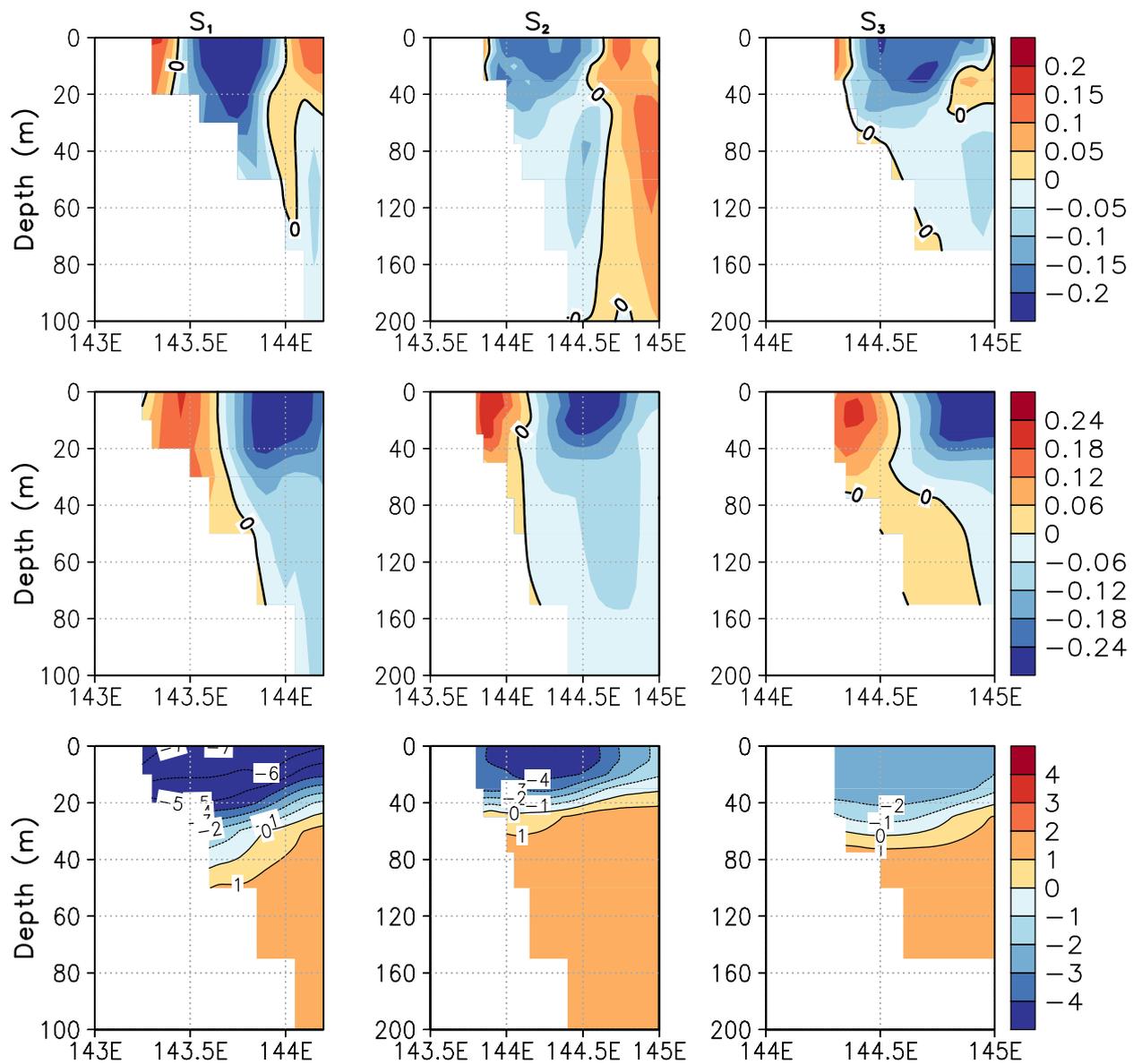}
\caption{Vertical structure on the zonal sections:
(143$^{\circ}$E--144.2$^{\circ}$E, 52$^{\circ}$N) left column,
(143.5$^{\circ}$E--145$^{\circ}$E, 50.46$^{\circ}$N) central column
and (144$^{\circ}$E--145$^{\circ}$E, 49.51$^{\circ}$N) right column
across the eastern shelf of Sakhalin Island on 8 April 2005: (a)
vertical component of relative vorticity (shading), (b) meridional
velocity (shading, m s$^{-1}$) and (c) density deviation (shading,
kg m$^{-3}$) from the reference value $\rho_0$ amounts to 1025 kg
m$^{-3}$} \label{fig:11}
\end{figure}

According to the presented estimations, the considered variables
reach their maximal magnitudes in winter. The time-varying wind
stress component plays the leading role in the generation of EKE and
its contribution amounts to about 4 GW. The intensity of energy
conversion from APE to EPE amounts to 0.9 GW, whereas the intensity
of energy conversion from EPE to EKE amounts to 0.3 GW. Negative
sign of the integrated \textit{BT} indicates predominance of energy
conversion from EKE to MKE and its intensity amounts to
4.2$\times$10$^{-3}$ GW. It is induced by strong heterogeneous
spatial distribution of the \textit{BT} (see, fig.~\ref{fig:6}).
Minimal magnitudes for all considered variables are observed in
summer.

Thus, major sources of the EKE along the western boundary of the Sea
of Okhotsk are the time-varying wind stress and baroclinic
instability of the along-shore branch of the ESC, characterized by
strong seasonal variability. The significance of the baroclinic
instability of the ESC in the generation of the EKE coincides with
conclusions about the leading role of baroclinic instability in the
generation of mesoscale variability both in the World Ocean
\cite{Stammer1997,Storch2012} and other seas
\cite{Thomson1984,Yang2013}. Wind power input, namely, its
time-varying component also plays significant role in the generation
of the EKE both in the World Ocean \cite{Huang2006} and other basins
\cite{Yang2013}.

To demonstrate the result of the baroclinic instability of the ESC,
we have examined the horizontal velocity filed and vertical
component of the relative vorticity vector on the eastern shelf of
Sakhalin Island during winter-spring period 2005. The analysis of
these fields has revealed eddy structures generated on the eastern
shelf of Sakhalin Island from March to April 2005. Mean spatial
scale of these eddies and the first baroclinic Rossby radius of
deformation are the same order and mean lifetime of the eddies
amounts to about 45 days.

Thus, these revealed eddies are the mesoscale eddies induced by
baroclinic instability of the ESC. These mesoscale eddies induce
eddy buoyancy flux $-\overline{\rho'w'}g$, which is very similar the
vertical eddy heat flux \cite{Wolfe2008}. High vertical eddy
buoyancy flux will result in strong vertical mixing on the eastern
shelf of Sakhalin Island. These revealed mesoscale variability needs
taking into account at analyzing the intermediate water transport
with the ESC. In addition, at forecasting the evolution of the
ecosystem on the eastern shelf of Sakhalin Island also needs take
into account mesoscale variability during winter-spring period.

\begin{acknowledgements}
This work was supported by the RFBR (project 17-05-00035) and by the
POI FEBRAS Program 'Mathematical simulation and analysis of
dynamical processes in the ocean' (number 117030110034-7).
\end{acknowledgements}

\bibliographystyle{spmpsci}
\bibliography{EKE_OKH}   

\newpage
FIGURE AND TABLE CAPTIONS

{\bf Fig.~1} Annual mean first baroclinic Rossby radius of
deformation in the Sea of Okhotsk (shading, km). Bottom topography
of the Sea of Okhotsk extracted from the GEBCO dataset and smoothed
by 9--point filter (lines, m). The rectangle marks out the region
near the eastern Sakhalin Island (141.6$^{\circ}$E--146$^{\circ}$E,
44$^{\circ}$N--55$^{\circ}$N).

{\bf Fig.~2} Seasonal mean wind power input (shading, 10$^{-2}$W
m$^{-2}$) and velocity field at the horizon of 10 m (vectors, m
s$^{-1}$) in (a) winter (January, February, March), (b) spring
(April, May, June), (c) summer (July, August, September) and (d)
autumn (October, November, December).

{\bf Fig.~3} Basin-averaged vertical profiles of the EKE (red line)
and MKE (blue line) over the Sea of Okhotsk in winter (January,
February, and March) (dashed line) and summer (July, August and
September) (solid line).

{\bf Fig.~4} The mean EKE (10$^{3}$ J m$^{-2}$) integrated in the
upper 200 m in (a) winter, (b) spring, (c) summer and (d) autumn.

{\bf Fig.~5} Distribution of the rate of the energy conversion term
(\textit{BC}) (10$^{-2}$ W m$^{-2}$) integrated in the upper 200 m
in (a) winter, (b) spring, (c) summer and (d) autumn.

{\bf Fig.~6} Distribution of the rate of the energy conversion term
(\textit{BT}) (10$^{-4}$ W m$^{-2}$) integrated in the upper 200 m
in (a) winter, (b) spring, (c) summer and (d) autumn.

{\bf Fig.~7} Distribution of generation of EKE due to time-varying
wind stress (10$^{-2}$ W m$^{-2}$) in (a) winter, (b) spring, (c)
summer and (d) autumn.

{\bf Fig.~8} Distribution of the rate of the energy conversion term
$(-\overline{\rho'w'}g)$ (10$^{-3}$ W m$^{-2}$) integrated in the
upper 200 m in (a) winter, (b) spring, (c) summer and (d) autumn.

{\bf Fig.~9} Vertical section of monthly mean meridional velocity
(shedding, m s$^{-1}$) and density deviation (lines, kg m$^{-3}$)
from the reference value $\rho_0$, amounting to 1025 kg m$^{-3}$,
across the eastern shelf of Sakhalin Island (50.46$^{\circ}$N) in
(a) January, (b) February and (c) March 2005. Left-hand side (lines,
10$^{-3}$ s$^{-1}$) and right-hand side (shading, 10$^{-3}$
s$^{-1}$) of relation (15) in: (d) January, (e) February, and (f)
March 2005.

{\bf Fig.~10} Velocity field (vectors, m s$^{-1}$) and vertical
component of relative vorticity field (shading, 10$^{-1}$) at the
horizon of 20 m on the eastern shelf of Sakhalin Island on 8 April
2005.

{\bf Fig.~11} Vertical structure on zonal sections:
(143$^{\circ}$E--144.2$^{\circ}$E, 52$^{\circ}$N) left column,
(143.5$^{\circ}$E--145$^{\circ}$E, 50.46$^{\circ}$N) central column
and (144$^{\circ}$E--145$^{\circ}$E, 49.51$^{\circ}$N) right column
across the eastern shelf of Sakhalin Island on 8 April 2005: (a)
vertical component of relative vorticity (shading), (b) meridional
velocity (shading, m s$^{-1}$) and (c) density deviation (shading,
kg m$^{-3}$) from the reference value $\rho_0$ amounts to 1025 kg
m$^{-3}$.

{\bf Table~1} Long-term mean rates of energy conversion (\textit{BT}
and \textit{BC}) and magnitudes of two sources of the EKE
($\overline{\mathbf{\tau}' \cdot \mathbf{u}'_{s}}$ and
$-\overline{\rho' w'g}$), integrated in the upper 200 m on the
eastern shelf of Sakhalin Island (141.6$^{\circ}$E--146$^{\circ}$E,
44$^{\circ}$N--55$N^{\circ}$). Unit is in 10$^{9}$W.
\newpage
%
%
%
\newpage
%
%
\newpage
%
%
\newpage
%
%
%
\newpage
%
%
%
\newpage
%
%
%
\newpage
%
%
%
\newpage
%
%
%
\newpage
%
%
%
\newpage
%
%
%
\newpage
%
%
%
%
%
\clearpage
\begin{table}[h]
\caption{Long-term mean rates of energy conversion (\textit{BT} and
\textit{BC}) and magnitudes of two sources of the EKE
($\overline{\mathbf{\tau}' \cdot \mathbf{u}'_{s}}$ and
$-\overline{\rho' w'g}$), integrated in the upper 200 m on the
eastern shelf of Sakhalin Island (141.6$^{\circ}$E--146$^{\circ}$E,
44$^{\circ}$N--55$^{\circ}$N). Unit is in 10$^{9}$W} \centering
\label{tab:1}
\begin{tabular}{l ccccc}
\hline\noalign{\smallskip}
 Season     & $BT=-\rho_0\overline{\mathbf{u}'_{h} \cdot (\mathbf{u}'\cdot \nabla \overline{\mathbf{u}_h}) }$ & $BC=-\frac{g^2}{\overline{N}^2\rho_0}\overline{\mathbf{u}'_h\rho'}\cdot \nabla_{h} \overline{\rho}$ & $\overline{\mathbf{\tau}' \cdot
\mathbf{u}'_{s}}$ & $-\overline{\rho' w'g}$\\
\hline\noalign{\smallskip}
  $Winter$  & $-4.2\cdot10^{-3}$ & 0.9 & 4.0 & 0.3  \\
  $Spring$  & $-2.7\cdot10^{-3}$ & 0.1 & 1.2 & $2\cdot10^{-2}$\\
  $Summer$  & $-1.9\cdot10^{-3}$ & 0.1 & 0.9 & $-2.9\cdot10^{-2}$\\
  $Autumn$  & $-2.1\cdot10^{-3}$ & 0.5 & 2.9 & $5.8\cdot10^{-2}$\\
\hline\noalign{\smallskip}
\end{tabular}
\end{table}

\end{document}